\documentclass[12pt]{elsart}
\makeatletter\def\ps@copyright{}\makeatother

\usepackage[a4paper, height=220mm, width=155mm]{geometry}
\usepackage{amsmath, amssymb}
\usepackage{graphicx}

\newcommand{\eVq}{\ensuremath{\text{eV}^2}}
\newcommand{\Dmq}{\Delta m^2}
\newcommand{\Sol}{\text{sol}}
\newcommand{\Atm}{\text{atm}}
\newcommand{\Sbl}{\textsc{sbl}}

\begin{document}

\begin{frontmatter}

\begin{flushright}
    IFT-UAM/CSIC-07-16
\end{flushright}

\title{Sterile neutrinos at the CNGS} 

\author[ad:Madrid]{Andrea Donini,}
\author[ad:Madrid]{Michele Maltoni,}
\author[ad:Roma]{Davide Meloni,}
\author[ad:Napoli]{Pasquale Migliozzi,}
\author[ad:Frascati]{Francesco Terranova}

\address[ad:Madrid]{%
  Instituto F\'{\i}sica Te\'orica UAM/CSIC,
  Cantoblanco, E-28049 Madrid, Spain}

\address[ad:Roma]{%
  I.N.F.N., Sezione di Roma I and Dip.\ Fisica, Univ.\ Roma ``La
  Sapienza'', Pl.~A.~Moro 2, I-00185, Rome, Italy}

\address[ad:Napoli]{%
  I.N.F.N., Sezione di Napoli, I-80126, Naples, Italy}

\address[ad:Frascati]{%
  I.N.F.N., Laboratori Nazionali Frascati, Via E.~Fermi 40, I-00044,
  Frascati, Italy}

\begin{abstract}
    We study the potential of the CNGS beam in constraining the
    parameter space of a model with one sterile neutrino separated
    from three active ones by an $\mathcal{O}(\eVq)$ mass-squared
    difference, $\Dmq_\Sbl$.  We perform our analysis using the OPERA
    detector as a reference (our analysis can be upgraded including a
    detailed simulation of the ICARUS detector). We point out that the
    channel with the largest potential to constrain the sterile
    neutrino parameter space at the CNGS beam is $\nu_\mu \to
    \nu_\tau$. The reason for that is twofold: first, the
    active-sterile mixing angle that governs this oscillation is the
    less constrained by present experiments; second, this is the
    signal for which both OPERA and ICARUS have been designed, and
    thus benefits from an extremely low background. In our analysis we
    also took into account $\nu_\mu \to \nu_e$ oscillations.  We find
    that the CNGS potential to look for sterile neutrinos is limited
    with nominal intensity of the beam, but it is significantly
    enhanced with a factor 2 to 10 increase in the neutrino flux. Data
    from both channels allow us, in this case, to constrain further
    the four-neutrino model parameter space. Our results hold for any
    value of $\Dmq_\Sbl \gtrsim 0.1~\eVq$, \textit{i.e.}\ when
    oscillations driven by this mass-squared difference are averaged.
    We have also checked that the bound on $\theta_{13}$ that can be put
    at the CNGS is not affected by the possible existence of sterile neutrinos.
\end{abstract}

\vspace*{\stretch{2}}
\begin{flushleft}
\end{flushleft}

\end{frontmatter}

\section{Introduction}

The results of solar~\cite{Cleveland:1998nv, Abdurashitov:1999zd,
Hampel:1998xg, Fukuda:2001nj, Ahmad:2001an, Ahmed:2003kj},
atmospheric~\cite{Fukuda:1998mi, Ambrosio:2001je},
reactor~\cite{Apollonio:1999ae, Apollonio:2002gd, Boehm:2001ik,
Eguchi:2002dm} and accelerator~\cite{Ahn:2002up, Aliu:2004sq,
Michael:2006rx} neutrino experiments show that flavour mixing occurs
not only in the hadronic sector, as it has been known for long, but in
the leptonic sector as well.  The full understanding of the leptonic
mixing matrix constitutes, together with the discrimination of the
Dirac/Majorana character of neutrinos and with the measurement of
their absolute mass scale, the main goal of neutrino physics for the
next decade. 

The experimental results point to two very distinct mass-squared
differences, $\Dmq_\Sol \approx 7.9 \times 10^{-5}~\eVq$ and
$|\Dmq_\Atm| \approx 2.4 \times 10^{-3}~\eVq$. On the other hand, only
two out of the four parameters of the three-family leptonic mixing
matrix $U_\text{PMNS}$~\cite{Pontecorvo:1957cp, Maki:1962mu,
Pontecorvo:1967fh, Gribov:1968kq} are known: $\theta_{12} \approx
34^\circ$ and $\theta_{23} \approx
43^\circ$~\cite{GonzalezGarcia:2007ib}.  The other two parameters,
$\theta_{13}$ and $\delta$, are still unknown: for the mixing angle
$\theta_{13}$ direct searches at reactors~\cite{Apollonio:1999ae,
Apollonio:2002gd, Boehm:2001ik} and three-family global analysis of
the experimental data give the upper bound $\theta_{13} \leq
11.5^\circ$, whereas for the leptonic CP-violating phase $\delta$ we
have no information whatsoever (see, however,
Ref.~\cite{GonzalezGarcia:2007ib}).

The LSND data~\cite{Athanassopoulos:1996wc, Athanassopoulos:1997pv,
Aguilar:2001ty}, on the other hand, would indicate a $\bar \nu_\mu \to
\bar \nu_e$ oscillation with a third neutrino mass-squared difference:
$\Dmq_\text{LSND} \sim 0.3 - 6~\eVq$, about two orders of magnitude
larger than $\Dmq_\Atm$. Given the strong hierarchy among the solar,
atmospheric and LSND mass-squared splittings, $\Dmq_\Sol \ll \Dmq_\Atm
\ll \Dmq_\Sbl$, it is not possible to explain all these data with just
three massive neutrinos, as it has been shown with detailed
calculations in Ref.~\cite{Fogli:1999zq}. A necessary condition to
explain the whole ensemble of data in terms of neutrino oscillations
is therefore the introduction of \emph{at least} a fourth light
neutrino state.  This new light neutrino must be an electroweak
singlet~\cite{Pontecorvo:1967fh} in order to comply with the strong
bounds on the $Z^0$ invisible decay width~\cite{Yao:2006px, LEPfinal}.
For this reason, the LSND signal has often been considered as an
evidence of the existence of a sterile neutrino.

In recent years, global analyses of solar, atmospheric,
short-baseline~\cite{Kleinfeller:2000em, Dydak:1983zq,
Stockdale:1984ce, Declais:1994su} experiments and LSND data have been
performed to establish whether four-neutrino models can really
reconcile the data and solve the puzzle~\cite{Grimus:2001mn,
Bilenky:1996rw, Okada:1996kw, Barger:1998bn, Bilenky:1999ny,
Peres:2000ic, Giunti:2000ur, Maltoni:2004ei}. The point is that
providing a suitable mass-squared difference to each class of
experiments is not enough: it is also necessary to show that the
intrinsic structure of the neutrino mixing matrix is compatible with
all the data. This turned out to be very hard to accomplish. In
Ref.~\cite{Maltoni:2002xd} it was shown that four-neutrino models were
only marginally allowed, with best fit around $\Dmq_\text{LSND} \simeq
1~\eVq$ and $\sin^2 2 \theta_\text{LSND} \simeq 10^{-3}$. Generically
speaking, the global analysis indicated that a single sterile neutrino
state was not enough to reconcile LSND with the other experiments. For
this reason, to improve the statistical compatibility between the LSND
results and the rest of the oscillation data, models with two sterile
neutrino states have been tested (see, for example,
Ref.~\cite{Sorel:2003hf} and references therein). Although a slightly
better global fit was achieved, a strong tension between the LSND data
and the results from atmospheric and short-baseline experiments was
still present. 

So far, the LSND signal has not been confirmed by any other
experiment~\cite{Church:2002tc}. It is therefore possible that the
LSND anomaly arises from some some yet unknown problem in the data set
itself. To close the issue, the MiniBooNE
collaboration~\cite{AguilarArevalo:2007it} at FermiLab has recently
performed a search for $\nu_\mu \to \nu_e$ appearance with a baseline
of 540 m and a mean neutrino energy of about 700 MeV. The primary
purpose of this experiment was to test the evidence for $\bar \nu_\mu
\to \bar \nu_e$ oscillation observed at LSND with a very similar $L/E$
range.  No evidence of the expected signal has been found, hence
ruling out once and for all the four-neutrino interpretation of the
LSND anomaly. However, MiniBooNE data are themselves not conclusive:
although no evidence for $\nu_\mu \to \nu_e$ oscillation has been
reported in the spectrum region compatible with LSND results, an
unexplained excess has been observed for lower energy neutrinos.
Furthermore, within a five-neutrino model this excess can be easily
explained, and even reconciled with LSND and all the other
\emph{appearance} experiments~\cite{Maltoni:2007zf}. On the other
hand, a post-MiniBooNE global analysis including also
\emph{disappearance} data show that five-neutrino models suffer from
the same problems as four-neutrino schemes, and in particular they are
now only marginally allowed -- a situation very similar to that of
four-neutrino models before MiniBooNE data. Adding a third sterile
neutrino\footnote{A quite interesting scenario is, in our opinion,
that in which three right-handed Majorana neutrinos are added to the
three weakly-interacting ones.  If the Majorana mass term $M$ is
$\mathcal{O}(\text{eV})$, (3+3) light Majorana neutrinos are present
at low-energy~\cite{deGouvea:2005er, deGouvea:2006gz}.} does not
help~\cite{Maltoni:2007zf}, and in general global analyses seem to
indicate that sterile neutrinos alone are not enough to reconcile all
the data. Models with sterile neutrinos and exotic physics have been
therefore proposed (see, for example,
Ref.~\cite{PalomaresRuiz:2005vf}).

In summary, the present experimental situation is still confused. It
is therefore worthwile to understand if, aside of MiniBooNE, new
neutrino experiments currently running or under construction can
investigate the existence of sterile neutrinos separated from the
active ones by $\mathcal{O}(\eVq)$ mass-squared differences. In this
paper we explore in detail the capability of the CNGS beam to perform
this search. For definiteness we focus on the simplest case with only
one extra sterile neutrino. Note that this model is perfectly viable
once the LSND result is dropped, as it contains as a limiting case the
usual three-neutrino scenario. Furthermore, it is easily generalizable
by adding new sterile neutrino states, and it can be used as a basis
for models with extra ``sterile'' states strongly decoupled from
active neutrinos (such as in extra-dimensions models with a
right-handed neutrino in the bulk~\cite{Pas:2005rb}).

The CNGS beam~\cite{Giacomelli:2007df} has been built to test the
(supposedly) dominant oscillation in atmospheric neutrino data,
$\nu_\mu \to \nu_\tau$.  In order to make possible $\tau$ production
through CC interactions, the mean neutrino energy, $\left< E_\nu
\right> = 17$ GeV, is much above the atmospheric oscillation peak for
the CERN to Gran Sasso baseline, $L = 732$ Km. Two detectors are
illuminated by the CNGS beam: OPERA (see
Ref.~\cite{Acquafredda:2006ki} and refs. therein) will start data
taking with the lead-emulsion target in 2007; ICARUS-T600 (see
Ref.~\cite{Amerio:2004ze} and refs. therein) will start operating in
2008. Both detectors have been especially designed to look for
$\tau$'s produced through $\nu_\mu \to \nu_\tau$ oscillation and to
minimize the corresponding backgrounds. The expected number of $\tau$
events after signal selection in an experiment such as OPERA (after
five years of data taking with nominal CNGS luminosity) is
$\mathcal{O}(10)$ events with $\mathcal{O}(1)$ background event. 

At the CNGS distance and energy, neutrino oscillations mediated by an
$\mathcal{O}(\eVq)$ mass difference will appear as a constant term in
the oscillation probability. In four-neutrino models, fluctuations
induced by this term over the atmospheric $\nu_\mu \to \nu_\tau$
oscillation can be as large as 100\% for specific points of the
allowed parameter space.  This is due to the fact that the leading
angle for this oscillation is the less constrained one. 
The $\nu_\mu \to \nu_\tau$ channel, therefore, is extremely promising
as a ``sterile neutrino'' smoking gun, as it has been commented
elsewhere (see, for example, Refs.~\cite{Donini:2001xy, Donini:2001xp}
and refs. therein).  To test the model we will also make use of the
$\nu_\mu \to \nu_e$ channel. Notice that the background to this signal
coming from $\tau \to e$ decay is modified in four-neutrino models
with respect to standard three-family oscillations. In fact, since
$\nu_\mu \to \nu_\tau$ oscillations are depleted by active-sterile
mixing with respect to standard ones, the $\tau \to e$ background to
$\nu_\mu \to \nu_e$ oscillations gets depleted, too. A combined
analysis of the two channels in four-neutrino models at the OPERA
detector has been performed, taking into account properly all of the
backgrounds. We stress, however, that the same analysis could be
performed at ICARUS, as well. The previous considerations hold for any
facility operating well beyond the kinematical threshold for $\tau$
production.

In the specific case of the CNGS beam, the limited flux implies a
modest improvement in the parameter space exclusion, see
Sec.~\ref{sec:results}.  An increase in the exposure of such
facilities, however, would permit to improve the present bounds on the
parameters of four-neutrino models and, in particular, to constrain
the leading angle in $\nu_\mu \to \nu_\tau$ oscillations at the level
of the other mixing parameters. 

The paper is organized as follows. In Sec.~\ref{sec:schemes} we
briefly review the main features of four-neutrino models and we
introduce our parametrization of the mixing matrix. In
Sec.~\ref{sec:probs} we compute the vacuum oscillation probabilities
in the atmospheric regime and we review the present bounds on the
active-sterile mixing angles. In Sec.~\ref{sec:CNGS} we recall the
most relevant parameters of CNGS. In Sec.~\ref{sec:app} we study
theoretically the expectations of the $\nu_\mu \to \nu_\tau$ and
$\nu_\mu \to \nu_e$ channels at the CNGS. In Sec.~\ref{sec:results} we
present our results using these channels at the OPERA detector and the
CNGS beam. Finally, in Sec.~\ref{sec:concl} we draw our conclusions.

\section{Four neutrino mass schemes}
\label{sec:schemes}

\begin{figure}[t] \centering
    \includegraphics[width=0.7\linewidth]{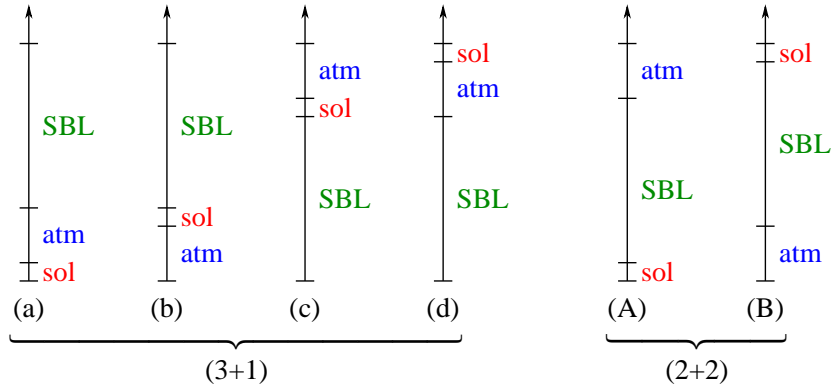}
    \caption{\label{fig:schemes}\sl%
      The two classes of four--neutrino mass spectra, (3+1) and
      (2+2).}
\end{figure}

In four-neutrino models, one extra sterile state is added to the three
weakly interacting ones. The relation between the flavor and the mass
eigenstates is then described by a $4 \times 4$ unitary matrix $U$,
which generalizes the usual $3 \times 3$ matrix
$U_\text{PMNS}$~\cite{Pontecorvo:1957cp, Maki:1962mu,
Pontecorvo:1967fh, Gribov:1968kq}.
As stated in the introduction, in this work we only consider the case
when the fourth mass eigenstate is separated by the other three by an
$\mathcal{O}(\eVq)$ mass-squared gap. There are six possible
four-neutrino schemes, shown in Fig.~\ref{fig:schemes}, that can
accommodate the results from solar and atmospheric neutrino
experiments and contain a third much larger $\Delta m^2$.  They can be
divided in two classes: (3+1) and (2+2). In the (3+1) schemes, there
is a group of three close-by neutrino masses that is separated from
the fourth one by the larger gap.  In (2+2) schemes, there are two
pairs of close masses separated by the large gap. While different
schemes within the same class are presently indistinguishable, schemes
belonging to different classes lead to very different phenomenological
scenarios.

A characteristic feature of (2+2) schemes is that the extra sterile
state cannot be simultaneously decoupled from \emph{both} solar and
atmospheric oscillations. To understand why, let us define
\begin{equation}
    \eta_s = \sum_{i \,\in\, \text{sol}} |U_{s i}|^2
    \qquad\text{and}\qquad
    c_s = \sum_{j \,\in\, \text{atm}} |U_{s j}|^2
\end{equation}
where the sums in $i$ and $j$ run over mass eigenstates involved in
solar and atmospheric neutrino oscillations, respectively. Clearly,
the quantities $\eta_s$ and $c_s$ describe the fraction of sterile
neutrino relevant for each class of experiment. 
Results from atmospheric and solar neutrino data imply that in both
kind of experiments oscillation takes place mainly between active
neutrinos. Specifically, from Fig.~46 of
Ref.~\cite{GonzalezGarcia:2007ib} we get $\eta_s \le 0.30$ and $c_s
\le 0.36$ at the $3\sigma$ level. However, in (2+2) schemes unitarity
implies $\eta_s + c_s = 1$, as can be easily understood by looking at
Fig.~\ref{fig:schemes}. These models are therefore ruled out at a very
high confidence level~\cite{Maltoni:2002ni}, and in the rest of this
work we will not consider them anymore.

On the other hand, (3+1) schemes are not affected by this problem.
Although the experimental bounds on $\eta_s$ and $c_s$ quoted above
still hold, the condition $\eta_s + c_s = 1$ no longer applies. For
what concerns neutrino oscillations, (3+1) models are essentially
unfalsifiable, since they reduce to the conventional three-neutrino
scenario when the mixing between active and sterile states are small
enough.

The mixing matrix $U$ can be conveniently parametrized in terms of six
independent rotation angles $\theta_{ij}$ and three (if neutrinos are
Dirac fermions) or six (if neutrinos are Majorana fermions) phases
$\delta_i$.  In oscillation experiments, only the so-called ``Dirac
phases'' can be measured, the effect of the ``Majorana phases'' being
suppressed by factors of $m_\nu / E_\nu$. The Majorana or Dirac nature
of neutrinos can thus be tested only in $\Delta L = 2$ transitions
such as neutrino-less double $\beta$-decay~\cite{Bilenky:2001xq} or
lepton number violating decays~\cite{Yao:2006px}. In the following
analysis, with no loss in generality, we will restrict ourselves to
the case of 4 Dirac-type neutrinos only.

A generic rotation in a four-dimensional space can be obtained by
performing six different rotations along the Euler axes. Since the
ordering of the rotation matrices $R_{ij}$ (where $ij$ refers to the
plane in which the rotation takes place) is arbitrary, plenty of
different parametrizations of the mixing matrix $U$ are allowed. The
large parameter space (6 angles and 3 phases, to be compared with the
standard three-family mixing case of 3 angles and 1 phase) is however
reduced to a subspace whenever some of the mass differences become
negligible.  If the eigenstates $i$ and $j$ are degenerate, rotations
in the $ij$-plane become unphysical and the corresponding mixing angle
should drop from oscillation probabilities. If the matrix $R_{ij}$ is
the rightmost one the angle $\theta_{ij}$ automatically disappears,
since the matrix commutes with the vacuum hamiltonian.  The parameter
space gets therefore reduced to the physical angles and phases.  If a
different ordering of the rotation matrices is taken, no angle
explicitly disappears from the oscillation formulas, but the physical
parameter space is still smaller than the original one. In this case,
a parameter redefinition is needed to reduce the parameter space to
the observable sector.  In Refs.~\cite{Donini:1999jc, Donini:1999he}
it was shown how the one-mass dominance ($\Delta_\Sol \to 0$ and
$\Delta_\Atm \to 0$, where $\Delta = \Dmq L / 4
E$~\cite{DeRujula:1979yy}) and two-mass dominance ($\Delta_\Sol \to
0$) approximations can be implemented in a transparent way (in the
sense that only the physical parameters appear in oscillation
probabilities) using a parametrization in which rotations are
performed in the planes corresponding to smallest mass difference
first:
\begin{equation}
    \label{eq:3+1param}
    U_\text{SBL} =
    R_{14}(\theta_{14}) \; R_{24}(\theta_{24}) \; R_{34}(\theta_{34}) \; 
    R_{23}(\theta_{23} ,\, \delta_3) \; R_{13}(\theta_{13} ,\, \delta_2) \;
    R_{12}(\theta_{12} ,\, \delta_1) \,.
\end{equation}
This parametrization was shown to be particularly useful when
maximizing oscillations driven by a $\mathcal{O}(\eVq)$ mass
difference. The analytical expressions for the oscillation
probabilities in the (3+1) model in the one-mass dominance
approximation in this parametrization have been presented in
Ref.~\cite{Donini:2001xy}. 

In this paper, however, we are interested in a totally different
regime: the ``atmospheric regime'', with oscillations driven by the
atmospheric mass difference, $ \Dmq_\Atm L/ E \sim \pi/2$.  We will
then make use of the following parametrization, adopted in
Ref.~\cite{Maltoni:2007zf}:
\begin{equation}
    \label{eq:3+1param2}
    U_\text{atm} =
    R_{34}(\theta_{34}) \; R_{24}(\theta_{24}) \;
    R_{23}(\theta_{23} ,\, \delta_3) \;
    R_{14}(\theta_{14}) \; R_{13}(\theta_{13} ,\, \delta_2) \; 
    R_{12}(\theta_{12} ,\, \delta_1) \,.
\end{equation}
It is convenient to put phases in $R_{12}$ (so that it automatically
drops in the two-mass dominance regime) and $R_{13}$ (so that it
reduces to the ``standard'' three-family Dirac phase when sterile
neutrinos are decoupled). The third phase can be put anywhere; we will
place it in $R_{23}$.  Note that in the one-mass dominance regime all
the phases disappear from the oscillation probabilities.

\begin{figure}[t] \centering
    \includegraphics[width=0.95\textwidth]{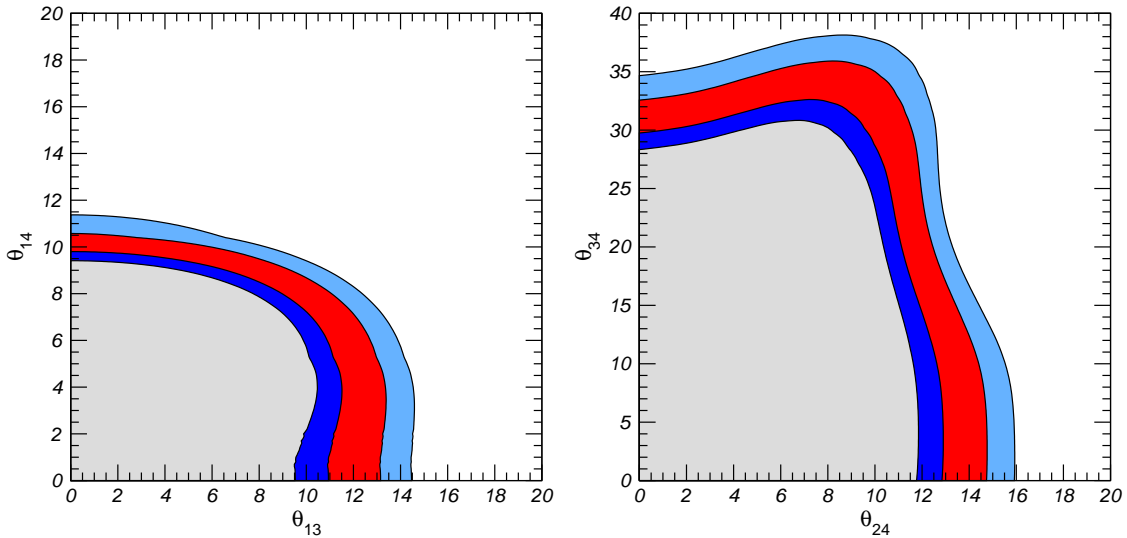}
    \caption{\label{fig:present}\sl%
      Allowed regions at 90\%, 95\%, 99\% and 3$\sigma$ CL in the
      $(\theta_{13},\, \theta_{14})$ plane (left) and in the
      $(\theta_{24},\, \theta_{34})$ plane (right) from the results of
      present atmospheric, reactor and LBL neutrino experiments. The
      undisplayed parameters $\theta_{23}$ and $\delta_3$ are
      marginalized.}
\end{figure}

\section{Oscillation probabilities and allowed parameter space}
\label{sec:probs}

Let us first consider $\nu_e$ disappearance at $L/E$ such that
$\Delta_\Sol$ can be safely neglected with respect to $\Delta_\Atm$
and $\Delta_\Sbl$. We get for this probability in vacuum:
\begin{equation}
    \label{eq:pee}
    P_{ee} = 1 - \sin^2 2 \theta_{14} \sin^2 \frac{\Delta_\Sbl L }{2} 
    - c^4_{14} \sin^2 2 \theta_{13} \sin^2 \frac{\Delta_\Atm L}{2} \, ,
\end{equation}
where $c_{ij} = \cos \theta_{ij}$ and $s_{ij} = \sin \theta_{ij}$. It
is clear from Eq.~\eqref{eq:pee} that reactor experiments such as
Bugey and Chooz can put stringent bounds to $\theta_{13}$ and
$\theta_{14}$, in this parametrization. This is depicted in
Fig.~\ref{fig:present}(left), where 90\%, 95\%, 99\% and 3$\sigma$ CL
contours in the ($\theta_{13}-\theta_{14}$)-plane are shown for
$\Delta_\Sol \to 0$ and $\Dmq_\Atm = 2.4 \times 10^{-3}~\eVq$. The
third mass difference, $\Dmq_\Sbl$, is free to vary above $0.1~\eVq$.
Notice that the $\nu_e$ disappearance probability does not depend on
$\theta_{23},\theta_{24}$ and $\theta_{34}$.  It can be clearly seen
that the three-family Chooz bound on $\theta_{13}$ is slightly
modulated by $\theta_{14}$.  Both angles, however, cannot be much
larger than $10^\circ$. We will therefore expand in these two
parameters to deduce the other relevant oscillation probabilities. 

At the CNGS beam atmospheric oscillations are large, solar
oscillations can be neglected and $\mathcal{O}(\eVq)$ oscillations are
extremely fast and can be averaged.  It is useful to write down the
oscillation probability (in vacuum) at typical atmospheric $L/E$, in
the approximation $\Delta_\Sol \to 0$, $\Delta_\Sbl \to \infty$. In
this regime:
\begin{equation} \begin{split}
    P(\nu_\alpha \to \nu_\beta) &= \delta_{\alpha \beta}
    - 4 \Re \left[ U_{\alpha 3} U^\star_{\beta 3} \left(
    \delta_{\alpha\beta} - U^\star_{\alpha 3} U_{\beta 3}
    - U^\star_{\alpha 4} U_{\beta 4} \right) \right]
    \sin^2 \frac{\Delta_{23} L }{2}
    \\
    &- 2 \Re \left[ \delta_{\alpha\beta} U_{\alpha 4} U^\star_{\beta 4}
    - |U_{\alpha 4}|^2 |U_{\beta 4}|^2 \right]
    \pm 2 \Im \left[ U_{\alpha 3} U^\star_{\beta 3}
    U^\star_{\alpha 4} U_{\beta 4} \right] \sin \Delta_{23} L \,,
\end{split} \end{equation}
where $+$ stands for neutrinos and $-$ for antineutrinos,
respectively. Up to second order in $\theta_{13}$ and $\theta_{14}$ we
get for the $\nu_\mu$ disappearance oscillation probability:
\begin{equation} \begin{split}
    P_{\mu\mu} &= 1 - 2 c^2_{14} s^2_{24} (1 - c^2_{14} s^2_{24})
    - 4 \big\lbrace s^2_{23} c^2_{24} \left[ c^2_{24} (c^2_{23}
    - s^2_{13}) - s^2_{14} s^2_{24} \right]
    \\    
    &- 2 c^3_{24} s_{23} (1 - 2 s^2_{23}) s_{13} s_{14} s_{24}
    \cos(\delta_2 - \delta_3) \big\rbrace
    \sin^2 \frac{\Delta_\Atm L }{2} \,.
\end{split} \end{equation}
A ``negative'' result in an atmospheric $L/E$ $\nu_\mu$ disappearance
experiment (such as, for example, K2K), in which $\nu_\mu$
oscillations can be very well fitted in terms of three-family
oscillations, will put a stringent bound on the mixing angle
$\theta_{24}$.  The bound from such experiments on $\theta_{24}$ can
be seen in Fig.~\ref{fig:present}(right), where 90\%, 95\%, 99\% and
3$\sigma$ CL contours in the ($\theta_{24}-\theta_{34}$)-plane are
shown for $\Delta_\Sol \to 0$ and $\Dmq_\Atm = 2.4 \times
10^{-3}~\eVq$.  The third mass difference, $\Dmq_\Sbl$, is free to
vary above $0.1~\eVq$. The mixing angles not shown have been fixed to:
$\theta_{23} = 45^\circ$; $\theta_{13} = \theta_{14} = 0$ (in this
hypothesis, $P_{\mu \mu}$ does not depend on phases).  Notice that the
$\nu_\mu$ disappearance probability does not depend on $\theta_{34}$. 

From the figure, we can see that $\theta_{24}$ cannot be much larger
than $10^\circ$, either. We will consider, therefore, the three mixing
angles $\theta_{13}, \theta_{14}$ and $\theta_{24}$ being of the same
order and expand in powers of the three.  At second order in
$\theta_{13}, \theta_{14}$ and $\theta_{24}$, we get:
\begin{equation}
    \label{eq:pmumu}
    P_{\mu\mu} = 1 - 2 s^2_{24}
    - 4 s^2_{23} \left[ c^2_{23} (1 - 2 s^2_{24}) - s^2_{13} \right]
    \sin^2 \frac{\Delta_\Atm L }{2} \,.
\end{equation}

Since both $\nu_e$ and $\nu_\mu$ disappearance do not depend on
$\theta_{34}$, we should ask which measurements give the upper bound
to this angle that can be observed in Fig.~\ref{fig:present}(right). 
This is indeed a result of indirect searches for $\nu_\mu \to \nu_s$
conversion in atmospheric experiments, using the different interaction
with matter of active and sterile neutrinos. This can be understood
from the (vacuum) $\nu_\mu \to \nu_s$ oscillation probability at
atmospheric $L/E$ for which, at second order in $\theta_{13},
\theta_{14}$ and $\theta_{24}$, we get:
\begin{equation} \begin{split}
    \label{eq:pmus}
    P_{\mu s} &= 2 c^2_{34} s^2_{24} + \big\lbrace
    \sin^2 2 \theta_{23} (c^4_{13} c^2_{24} s^2_{34}
    - c^2_{34} s^2_{24})
    \\
    &+ 2  c_{34} \sin 2 \theta_{23} s_{34} \left[
    s_{24} (1 -2 s^2_{23}) \cos \delta_3
    + 2 s_{23} s_{13} s_{14} \cos \delta_2 \right] \big\rbrace
    \sin^2 \frac{\Delta_\Atm L }{2}
    \\
    &\pm c_{34} \sin 2 \theta_{23} s_{24} s_{34}
    \sin\delta_3 \sin\Delta_\Atm L \,.
\end{split} \end{equation}

As it can be seen, the bound on $\theta_{34}$ arises from a
measurement of spectral distortion (\textit{i.e.}, from the
``atmospheric'' term proportional to $\sin^2 \Delta_\Atm L/2$). On the
other hand, bounds on $\theta_{13},\theta_{14}$ and $\theta_{24}$ are
mainly drawn by a flux normalization measurement. As a consequence,
the bound on $\theta_{34}$ that we can draw by non-observation of
$\nu_\mu \to \nu_s$ oscillation in atmospheric experiments is less
stringent than those we have shown before.  For this reason,
$\theta_{34}$ can be somewhat larger than $\theta_{13},\theta_{14}$
and $\theta_{24}$, but still bounded to be below $40^\circ$.  In the
following, we will expand in powers of the four mixing angles
$\theta_{13},\theta_{14}, \theta_{24}$ and $\theta_{34}$, that will be
considered to be comparably small. 

Up to fourth-order in $\theta_{13}, \theta_{14}, \theta_{24}$ and
$\theta_{34}$, the $\nu_\mu \to \nu_e$ appearance probability in the
atmospheric regime is: 
\begin{equation} \begin{split}
    \label{eq:pmue}
    P_{\mu e} &= 4 \left\lbrace s^2_{23} s^2_{13}
    [1 - s^2_{13} - s^2_{14} - s^2_{24}] 
    + s_{23} s_{13} s_{14} s_{24} \cos(\delta_2 - \delta_3) \right\rbrace
    \sin^2 \frac{\Delta_\Atm L}{2}
    \\
    & \pm 2 s_{23} s_{13} s_{14} s_{24} \sin(\delta_2 - \delta_3)
    \sin\Delta_\Atm L + 2 s^2_{14} s^2_{24} \,.
\end{split} \end{equation}

Eventually, the $\nu_\mu \to \nu_\tau$ appearance probability up to
fourth-order in $\theta_{13},\theta_{14},\theta_{24}$ and
$\theta_{34}$ in the atmospheric regime is: 
\begin{equation} \begin{split}
    \label{eq:pmutau}
    P_{\mu \tau} &= 2 s^2_{24} s^2_{34} + \big\lbrace \sin^2 2 \theta_{23}
    [c^4_{13} c^2_{24} c^2_{34} - s^2_{24} s^2_{34}]
    \\
    &- 4 \sin 2 \theta_{23} s_{13} s_{14} [s_{23} s_{34} \cos\delta_2
    + c_{23} s_{24} \cos(\delta_2 - \delta_3)]
    \\
    &+ 2 \sin 2 \theta_{23} s_{24} s_{34} c^2_{13} c^2_{24} c_{34}
    [c^2_{14} - 2 c^2_{13} s^2_{23}] \cos\delta_3 \big\rbrace
    \sin^2 \frac{\Delta_\Atm L}{2}
    \\
    &\mp \sin 2 \theta_{23} s_{24} s_{34} c^2_{13} c^2_{14} c^2_{24} c_{34}
    \sin\delta_3 \sin\Delta_\Atm L \,.
\end{split} \end{equation}

As it was shown in Refs.~\cite{Donini:2001xy, Donini:2001xp}, the
$\nu_\mu \to \nu_\tau$ appearance channel is a good place to look for
sterile neutrinos. This can be understood as follows: consider the
$\nu_\mu \to \nu_\tau$ three-family oscillation probability in the
atmospheric regime, up to fourth-order in $\theta_{13}$:
\begin{equation}
    P^{3\nu}_{\mu\tau} = P_{\mu\tau} (\theta_{i4} = 0)
    \simeq c^4_{13} \sin^2 2 \theta_{23}
    \sin^2 \frac{\Delta_\Atm L}{2} \,.
\end{equation}
The genuine active-sterile neutrino mixing effects are:
\begin{equation} \begin{split}
    \Delta P_{\mu \tau} &\equiv P_{\mu \tau} - P^{3\nu}_{\mu\tau}
    \\
    &= \left\lbrace - (s^2_{24} + s^2_{34}) \sin^2 2 \theta_{23} 
    + 2 \sin 2 \theta_{23} (1 - 2 s^2_{13}) s_{24} s_{34}
    \cos\delta_3 \right\rbrace \sin^2 \frac{\Delta_\Atm L}{2}
    \\
    &\mp \sin 2 \theta_{23} s_{24} s_{34} \sin\delta_3
    \sin \Delta_\Atm L + \dots
\end{split} \end{equation}
that is second-order in small angles $\theta_{13}, \theta_{14},
\theta_{24}$ and $\theta_{34}$. We would get a similar result for
$\nu_\mu$ disappearance, also. On the other hand, computing the
corresponding quantity in the $\nu_\mu \to \nu_e$ channel, we get:
\begin{equation} \begin{split}
    \label{eq:dpmue}
    \Delta P_{\mu e} &\equiv P_{\mu e} - P_{\mu e}(\theta_{i4} = 0)
    \\
    &= s_{23} s_{13} s_{14} s_{24} \cos(\delta_2 - \delta_3)
    \sin^2 \frac{\Delta_\Atm L}{2}
    \\
    &\pm 2 s_{23} s_{13} s_{14} s_{24}
    \sin(\delta_2 - \delta_3) \sin \Delta_\Atm L + \dots
\end{split} \end{equation}
that is third-order in the same parameters.

Notice, eventually, that all oscillation probabilities start with an
energy-independent term and are, therefore, non-vanishing for $L=0$, a
result of our assumption that $\Delta_\Sbl \to \infty$.  

\section{The CNGS facility}
\label{sec:CNGS}

The CNGS is a conventional neutrino beam in which neutrinos are
produced by the decay of secondary pions and kaons, obtained from
collisions of 400 GeV protons from the CERN-SPS onto a graphite
target. The resulting neutrinos are aimed to the underground Gran
Sasso Laboratory (LNGS), located at 730 Km from CERN. This facility
provided the first neutrinos in August 2006~\cite{Acquafredda:2006ki}.
Differently from other long baseline experiments, the neutrinos from
CNGS can be exploited to search directly for $\nu_\mu \to
\nu_\tau$ oscillations, since they have a mean energy well beyond the
kinematic threshold for $\tau$ production. Moreover, the prompt
$\nu_\tau$ contamination (mainly from $D_s$ decays) is negligible. The
expected $\nu_e$ contamination is also relatively small compared to
the dominant $\nu_\mu$ component, thus allowing for the search of
sub-dominant $\nu_\mu \to \nu_e$ oscillations through an excess of
$\nu_e$ CC events.

\begin{figure}[t] \centering
    \includegraphics[width=10cm]{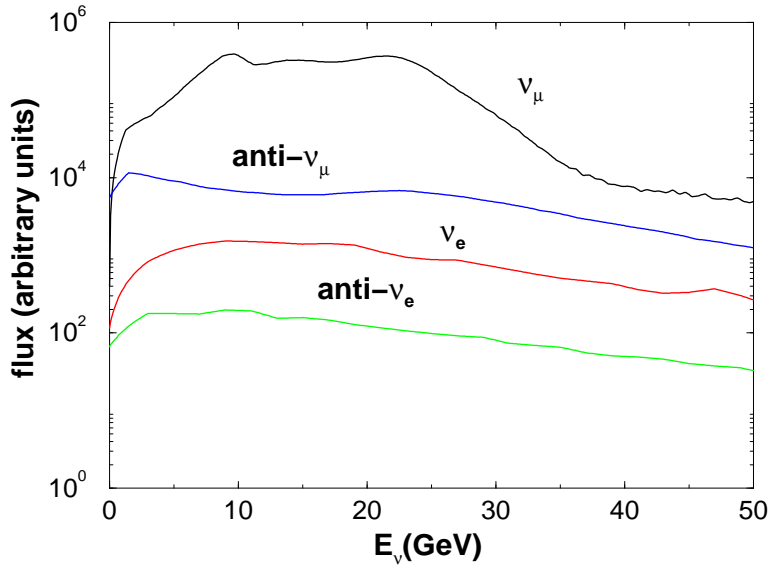}
    \caption{\label{fig:fluxes}\sl%
      CNGS neutrino fluxes (in arbitrary units) as a function of the
      neutrino energy.  Both muon and electron neutrino fluxes are
      illustrated.}
\end{figure}

The energy spectra of the CNGS neutrino beam are shown (in arbitrary
units) in Fig.~\ref{fig:fluxes}~\cite{fluxes}. In the present paper we
assume the nominal intensity for the CNGS, corresponding to $4.5
\times 10^{19}$ pot/year.

OPERA has been designed to search for $\tau$ appearance through
identification of the $\nu_\tau$ CC interaction on an event-by-event
basis. In particular, $\tau$'s are tagged identifying explicitly their
decay kink through high resolution nuclear emulsions interleaved with
lead sheets.  For this detector, we can take advantage of the detailed
studies of the $\nu_\mu \to \nu_\tau$ signal (see
Ref.~\cite{Migliozzi:2006fy}) and of the $\nu_\mu \to \nu_e$ signal
(see Ref.~\cite{Komatsu:2002sz}).

\begin{table}[t] \centering
    \begin{tabular}{|c|c|c|c|}
	\hline
	\hline
	$\nu_{\mu}$ &  $\bar \nu_{\mu}$ & $\nu_e$ &  $\bar \nu_e$\\
	\hline
	669.0 &13.7 & 5.9& 0.3\\
	\hline\hline
    \end{tabular}
    \caption{\label{tab:rates}\sl%
      Nominal performance of the CNGS reference beam~\cite{fluxes}.
      The total non-oscillated CC event rates are calculated assuming
      $10^{19}$ pot and 1 Kton lead target mass.}
\end{table}

The total non-oscillated CC event rates for a 1 Kton lead target with
a neutrino flux normalized to $10^{19}$ pot are shown in
Tab.~\ref{tab:rates} and are evaluated according to
\begin{equation}
    N = \int \frac{d \phi_{\nu_\alpha}(E)}{dE} \,
    \sigma_{\nu_\alpha}(E) \, dE \,,
\end{equation}
in which $\phi_{\nu_\alpha}$ is the flux of the neutrino flavour
$\nu_\alpha$ and $\sigma_{\nu_\alpha}$ the corresponding cross section
on lead.

\section{Appearance channels at the CNGS}
\label{sec:app}

\subsection{$\nu_\mu \to \nu_\tau$ oscillations}
\label{sec:apptau}

Since the CNGS experiments have been designed to search for $\nu_\mu
\to \nu_\tau$ oscillation in the parameter region indicated by the
atmospheric neutrino data, we can take full advantage of them in order
to constrain (and, possibly, study) the four-family parameter space. 

The number of taus from $\nu_\mu \to \nu_\tau$ oscillations is given
by the convolution of the $\nu_\mu$ flux $ d \phi_{\nu_{\mu}}(E) / dE$
with the $\nu_\tau$ charged-current cross-section on lead, 
$\sigma_{\nu_\tau}^{CC}(E)$, weighted by the $\nu_\mu \to \nu_\tau$
oscillation probability, $P_{\mu \tau}(E)$, times the efficiency for
the OPERA detector, $\varepsilon_{\mu \tau}$:
\begin{equation}
    \label{eq:signaltau}
    N_{\mu \tau} = A \int \frac{ d\phi_{\nu_{\mu}}(E)}{dE} \,
    P_{\mu \tau}(E) \sigma_{\nu_\tau}^{CC}(E) \,
    \varepsilon_{\mu \tau} \, dE \,.
\end{equation}
$A$ is a normalization factor which takes into account the target mass
and the normalization of the $\nu_\mu$ flux in physical units. 
Specializing our analysis for the OPERA detector, we have considered
an overall efficiency $\varepsilon_{\mu \tau} \sim
13\%$,~\cite{Migliozzi:2006fy}. This efficiency takes into account
that OPERA is able to exploit several decay modes of the final state
$\tau$, using both so-called short and long decays.  

The dominant sources of background for the $\nu_\mu \to \nu_\tau$
signal are charm decays and hadronic reinteractions.  Both of them
only depend on the total neutrino flux and not on the oscillation
probabilities.  The OPERA experiment at the CNGS beam has been
designed precisely to measure this channel, and thus the corresponding
backgrounds are extremely low.

\begin{table}[t] \centering
    \begin{tabular}{|c|c|c|c|c|c|}
	\hline\hline
	$(\theta_{13}; \theta_{14}; \theta_{24}; \theta_{34})$ & $N_\tau$ & background & $(\theta_{13}; \theta_{14}; \theta_{24}; \theta_{34})$ & $N_\tau$ & background \\
	\hline
	$(5^\circ;  5^\circ;  5^\circ; 20^\circ)$    &     8.9    & 1.0 & $(10^\circ; 5^\circ;  5^\circ; 20^\circ)$    &     8.5    & 1.0 \\  
	$(5^\circ;  5^\circ;  5^\circ; 30^\circ)$    &     6.9    & 1.0 & $(10^\circ; 5^\circ;  5^\circ; 30^\circ)$    &     6.5    & 1.0 \\
	$(5^\circ;  5^\circ; 10^\circ; 20^\circ)$    &     8.3    & 1.0 & $(10^\circ; 5^\circ; 10^\circ; 20^\circ)$    &     7.9    & 1.0 \\
	$(5^\circ;  5^\circ; 10^\circ; 30^\circ)$    &    10.5    & 1.0 & $(10^\circ; 5^\circ; 10^\circ; 30^\circ)$    &    10.3    & 1.0 \\
	3 families                          &    15.1    & 1.0 &          3 families                          &    14.4    & 1.0 \\
	\hline \hline
    \end{tabular}
    \caption{\label{tab:rates3}\sl%
      Event rates and expected background for the $\nu_\mu \to
      \nu_\tau$ channel in the OPERA detector, for different values of
      $\theta_{14}, \theta_{24}$ and $\theta_{34}$ in the (3+1)
      scheme. The other unknown angle, $\theta_{13}$ has been fixed
      to: $\theta_{13} = 5^\circ, 10^\circ$. The CP-violating phases
      are: $\delta_1 = \delta_2 = 0$; $\delta_3 = 90^\circ$.  As a
      reference, the expected value in the case of standard
      three-family oscillation (\textit{i.e.}, for $\theta_{i4} = 0$) is shown
      for maximal CP-violating phase $\delta$. The rates are computed
      according to Eq.~\eqref{eq:signaltau}.}
\end{table}

In Tab.~\ref{tab:rates3} we report the expected number of $\tau$
events in the OPERA detector, according to Eq.~\eqref{eq:signaltau},
for different values of $\theta_{13}, \theta_{14}, \theta_{24}$ and
$\theta_{34}$. Input points have been chosen according to the allowed
regions in the parameter space shown in Sec.~\ref{sec:probs}. The
other parameters are: $\theta_{12} = 34^\circ; \theta_{23} = 45^\circ;
\Dmq_\Sol = 7.9 \times 10^{-5}~\eVq$; $\Dmq_\Atm = 2.4 \times
10^{-3}~\eVq$ and $\Dmq_\Sbl = 1~\eVq$ (all mass differences are taken
to be positive).  Eventually, phases have been fixed to: $\delta_1 =
\delta_2 = 0; \delta_3 = 90^\circ$. The expected background is also
shown. Rates refer to a flux normalized to $4.5 \times 10^{19}$
pot/year (the nominal intensity of the CNGS), an active lead target
mass of 1.8 Kton and 5 years of data taking. For comparison, we also
report the expected number of events in the usual 3-family scenario.

As it can be seen, in most part of the parameter space we expect a
significant depletion of the signal with respect to standard
three-neutrino oscillations. However, the difference between (3+1)
model $\nu_\mu \to \nu_\tau$ oscillations and standard ones is much
bigger than the expected background. A good signal/noise separation
can therefore be used to test the model.

\subsection{$\nu_\mu \to \nu_e$ oscillations}
\label{sec:appe}

The number of electrons from the $\nu_\mu \to \nu_e$ oscillation is
given by the convolution of the $\nu_\mu$ flux $ d \phi_{\nu_{\mu}}(E)
/ dE$ with the $\nu_e$ charged-current cross-section on lead, 
$\sigma_{\nu_{e}}^{CC}(E)$, weighted by the $\nu_\mu \to \nu_e$
oscillation probability, $P_{\mu e}(E)$, times the efficiency for the
OPERA detector, $\varepsilon_{\mu e} (E)$~\cite{Komatsu:2002sz}:
\begin{equation}
    \label{eq:signale}
    N_{\mu e} = A \int \frac{ d\phi_{\nu_{\mu}}(E)}{dE} \,
    P_{\mu e}(E) \sigma_{\nu_{e}}^{CC}(E) \,
    \varepsilon_{\mu e}(E) \, dE \,,
\end{equation}
where $A$ is defined as above.  The overall signal efficiency $
\varepsilon_{\mu e}$ is the convolution of the kinematic efficiency
$\varepsilon_{\mu e}^{kin}$ (that ranges from 60\% to 80\% for
neutrino energies between 5 to 20 GeV) and several (nearly
factorizable) contributions.  Among them, the most relevant are
trigger efficiencies, effects due to fiducial volume cuts, vertex and
brick finding efficiencies and the electron identification capability.
They result in a global constant factor $\varepsilon_{\mu e}^{fact}
\sim$ 48\%.

The dominant sources of background to the $\nu_\mu \to \nu_e$ signal
are, in order of importance:
\begin{enumerate}
  \item $\nu_e$ beam contamination;
    
  \item fake electrons due to $\pi^0$ decays from $\nu_\mu$ NC
    interactions;
    
  \item electrons produced through $\tau$ decay, where the $\tau$
    comes from $\nu_\mu \to \nu_\tau$ oscillations;

  \item CC $\nu_\mu$ events where the muon is lost and a track mimics
    an electron.
\end{enumerate}

Backgrounds (1), (2) and (4) depend very little on the oscillation
parameters. On the other hand, the $\tau \to e$ background depends
strongly on the active-sterile mixing angles. As we have seen in
Sec.~\ref{sec:apptau}, in the allowed region of the parameter space
$\nu_\mu \to \nu_\tau$ oscillations are significantly depleted with
respect to the standard three-neutrino ones. As a consequence, this
background gets depleted, too. 

\begin{table}[t] \centering
    \begin{tabular}{|c|c|c|c|c|c|}
	\hline\hline
	$(\theta_{13}; \theta_{14}; \theta_{24}; \theta_{34})$ & $N_e$ & $\nu_e^{CC}$ & $\nu_\mu^{NC}$ & $\tau \to e$ & $\nu_\mu^{CC}$  \\ \hline
	$(5^\circ; 5^\circ; 5^\circ; 20^\circ)$   &  3.5 & 19.4 & 5.3 & 2.8 & 0.9 \\
	$(5^\circ; 5^\circ; 5^\circ; 30^\circ)$   &  3.5 & 19.4 & 5.3 & 2.1 & 0.9 \\
	$(5^\circ; 5^\circ; 10^\circ; 20^\circ)$  &  2.4 & 19.4 & 5.3 & 2.3 & 0.9 \\
	$(5^\circ; 5^\circ; 10^\circ; 30^\circ)$  &  2.4 & 19.4 & 5.3 & 2.4 & 0.9 \\
	3 families                        &  3.7 & 19.7 & 5.3 & 4.6 & 0.9 \\
	\hline
	$(10^\circ; 5^\circ; 5^\circ; 20^\circ)$  & 10.6 & 19.4 & 5.3 & 2.7 & 0.9 \\
	$(10^\circ; 5^\circ; 5^\circ; 30^\circ)$  & 10.4 & 19.4 & 5.3 & 2.0 & 0.9 \\
	$(10^\circ; 5^\circ; 10^\circ; 20^\circ)$ &  8.8 & 19.4 & 5.3 & 2.2 & 0.9 \\
	$(10^\circ; 5^\circ; 10^\circ; 30^\circ)$ &  8.6 & 19.4 & 5.3 & 2.4 & 0.9 \\
	3 families                        & 15.1 & 19.7 & 5.3 & 4.8 & 0.9 \\
	\hline \hline
    \end{tabular}
    \caption{\label{tab:rates2}\sl%
      Event rates and expected background for the $\nu_\mu \to \nu_e$
      channel in the OPERA detector, for different values of 
      $\theta_{14}, \theta_{24}$ and $\theta_{34}$ in the (3+1)
      scheme. The other unknown angle, $\theta_{13}$, has been fixed
      to: $\theta_{13} = 5^\circ, 10^\circ$. The CP-violating phases
      are: $\delta_1 = \delta_2 = 0$; $\delta_3 = 90^\circ$.  As a
      reference, the expected value in the case of standard
      three-family oscillation(\textit{i.e.}, for $\theta_{i4} = 0$) is shown
      for maximal CP-violating phase $\delta$.  The rates are computed
      according to Eq.~\eqref{eq:signale}. Backgrounds have been
      computed following Ref.~\cite{Komatsu:2002sz}.}
\end{table}

In Tab.~\ref{tab:rates2} we report the expected number of electrons in
the OPERA detector, according to Eq.~\eqref{eq:signale}, for different
values of $\theta_{13}, \theta_{14}, \theta_{24}$ and $\theta_{34}$.
Input points have been chosen according to the allowed regions in the
parameter space shown in Sec.~\ref{sec:probs}. The other parameters
are: $\theta_{12} = 34^\circ; \theta_{23} = 45^\circ; \Dmq_\Sol = 7.9
\times 10^{-5}~\eVq$; $\Dmq_\Atm = 2.4 \times 10^{-3}~\eVq$ and
$\Dmq_\Sbl = 1~\eVq$ (all mass differences are taken to be positive).
Eventually, phases have been fixed to: $\delta_1 = \delta_2 = 0;
\delta_3 = 90^\circ$. Backgrounds have been computed accordingly to
Ref.~\cite{Komatsu:2002sz}.  Rates refer to a flux normalized to $4.5
\times 10^{19}$ pot/year (the nominal intensity of the CNGS), an
active lead target mass of 1.8 Kton and 5 years of data taking. For
comparison, we also report the expected number of events in the usual
3-family scenario.

As it can be seen from Tab.~\ref{tab:rates2}, the difference between
the (3+1) model and the standard three-neutrino oscillations are
smaller in this channel than in the $\nu_\mu \to \nu_\tau$ one.
Moreover, they linearly depends on $\theta_{13}$, as it is clear from
Eq.~\eqref{eq:dpmue}. For $\theta_{13} = 5^\circ$, this channel will
be of no help to test the allowed parameter space of the (3+1) model.
On the other hand, for $\theta_{13}$ saturating the Chooz-Bugey bound,
both $\nu_\mu \to \nu_\tau$ and $\nu_\mu \to \nu_e$ might cooperate.
However, notice that backgrounds to this signal are much larger than
the difference between (3+1) model and standard three-neutrino
oscillations for any value of $\theta_{13}$.

\section{Sensitivity to $(3+1)$ sterile neutrinos at OPERA}
\label{sec:results}

In this section we study the sensitivity to $\theta_{13}$ and to the
active-sterile mixing angles $\theta_{14},\theta_{24}$ and
$\theta_{34}$ at the CNGS beam, using both the $\nu_\mu \to \nu_\tau$
and $\nu_\mu \to \nu_e$ appearance channels at the OPERA detector. In
the rest of this section, the known three-family subspace angles have
been fixed to: $\theta_{12} = 34^\circ; \theta_{23} = 45^\circ$. The
mass differences have been fixed to: $\Dmq_\Sol = 7.9 \times
10^{-5}~\eVq$ and $\Dmq_\Atm = 2.4 \times 10^{-3}~\eVq$.  The
CP-violating phases $\delta_1$ and $\delta_2$ have been kept fixed to
$\delta_1 = \delta_2 = 0$. On the contrary, the CP-violating phase 
$\delta_3$ is fixed to two values: $\delta_3 = 0$ or $90^\circ$. 
Notice that this phase is still present in the oscillation
probabilities even when $\theta_{12}$ and $\theta_{13}$ vanish, see
Eq.~\eqref{eq:pmutau}. At atmospheric $L/E$, oscillations driven by an
$\mathcal{O}(\eVq)$ mass difference are averaged. We have checked that
our results apply for any value of $\Dmq_\Sbl \geq 0.1~\eVq$. 

\begin{figure}[t]
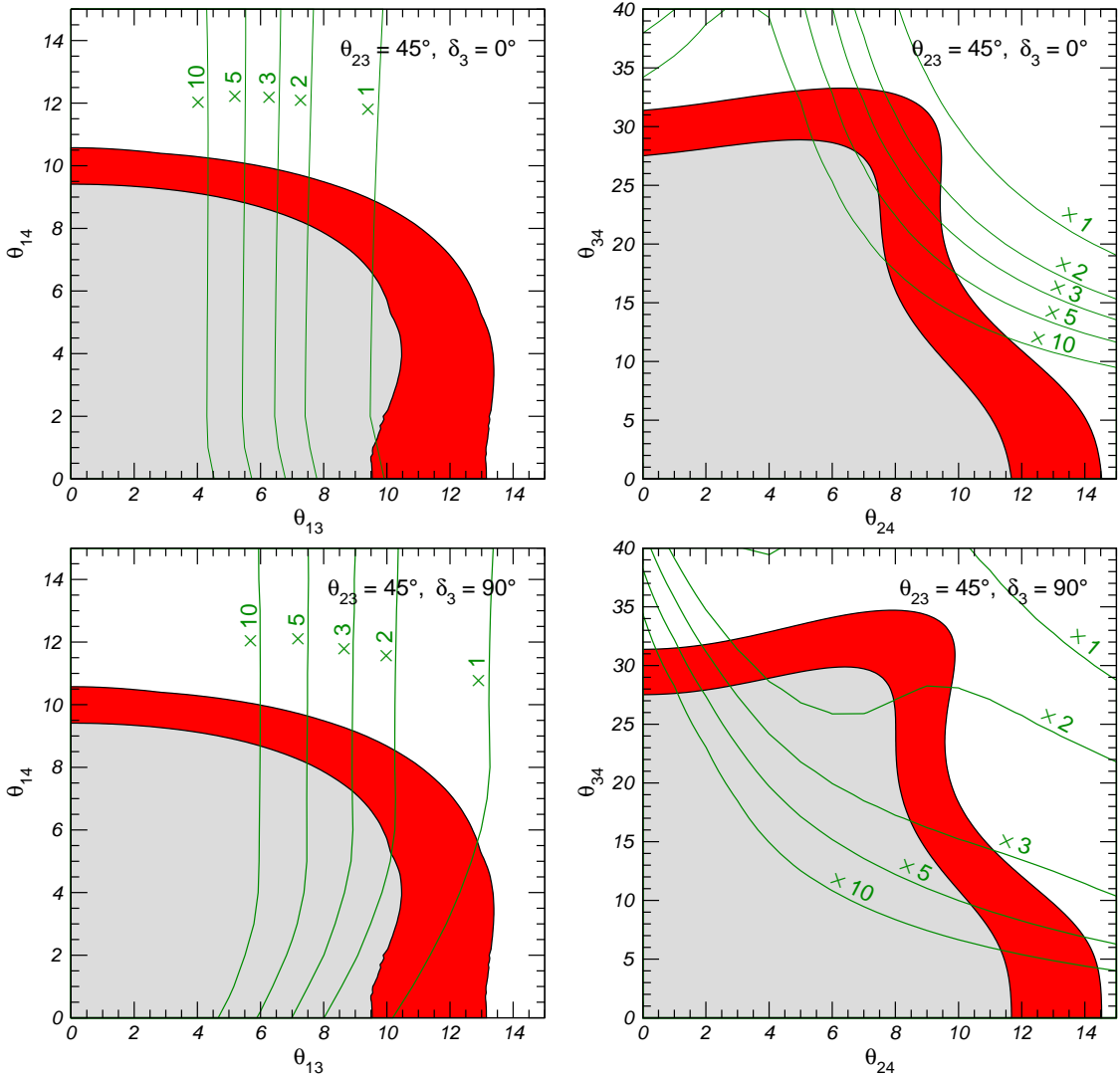
 \centering
    \includegraphics[width=0.95\textwidth]{fig.opera-00.eps}
    \includegraphics[width=0.95\textwidth]{fig.opera-90.eps}
    \caption{\label{fig:opera}\sl%
      Sensitivity limit at 99\% CL in the $(\theta_{13},\,
      \theta_{14})$ plane (left) and in the $(\theta_{24},\,
      \theta_{34})$ plane (right) from a null result of the OPERA
      experiment, assuming 1, 2, 3, 5 and 10 times the nominal
      intensity of $4.5 \times 10^{19}$ pot/year. The coloured regions
      show the present bounds at 90\% and 99\% CL. We assume
      $\theta_{23} = 45^\circ$ and $\delta_3 = 0^\circ$ (top) or
      $\delta_3 = 90^\circ$ (bottom).}
\end{figure}

In Fig.~\ref{fig:opera} we show the sensitivity limit at 99\% CL in
the $(\theta_{13},\,\theta_{14})$ plane (left) and in the
$(\theta_{24},\,\theta_{34})$ plane (right) from a null result of the
OPERA experiment, assuming 1, 2, 3, 5 and 10 times the nominal
intensity of $4.5 \times 10^{19}$ pot/year.  The coloured regions show
the present bounds at 90\% and 99\% CL. We assume $\theta_{23} =
45^\circ$ and $\delta_3 = 0^\circ$ (top) or $\delta_3 = 90^\circ$
(bottom). The sensitivity is defined as the region for which a
(poissonian) 2 d.o.f.'s $\chi^2$ is compatible with a ``null result''
at the 99\% CL.  We refer to ``null result'' when $\theta_{13}$ and
the three active-sterile mixing angles, $\theta_{14},\theta_{24}$ and
$\theta_{34}$ vanish simultaneously.  Both $\nu_\mu \to \nu_\tau$ and
$\nu_\mu \to \nu_e$ oscillations have been considered, with the
corresponding backgrounds treated properly as in Sec.~\ref{sec:app}. 
An overall systematic error of 10\% has been taken into account.  

In the left panels of Fig.~\ref{fig:opera} we can see that OPERA can
improve only a little the bound on $\theta_{13}$ after 5 years of data
taking working at nominal CNGS beam intensity, both for $\delta_3 = 0$
(top panel) or $\delta_3 = 90^\circ$ (bottom panel).  Increasing the
nominal intensity, however, a significant improvement on the bound is
achieved for any value of $\theta_{14}$. Notice that the limit on
$\theta_{14}$ is almost unaffected by the OPERA data.  This is because
for the $\nu_\mu \to \nu_\tau$ and $\nu_\mu \to \nu_e$ oscillation
probabilities at atmospheric $L/E$, the $\theta_{14}$-dependence
always arises at third-order in the small parameters
$\theta_{13},\theta_{14},\theta_{24}$ and $\theta_{34}$ (see
Eqs.~\eqref{eq:pmue} and~\eqref{eq:pmutau} for the explicit expression
in the adopted parametrization, Eq.~\eqref{eq:3+1param2}). On the
contrary, the $\theta_{13}$-, $\theta_{24}$- and
$\theta_{34}$-dependences in the same oscillation probabilities are
quadratic in the small parameters.  In case of vanishing
active-sterile mixing angles, $\theta_{i4} = 0$, see
Ref.~\cite{Komatsu:2002sz}. 

In the right panels of Fig.~\ref{fig:opera} the sensitivity of OPERA
to $\theta_{24}$ and $\theta_{34}$ is shown. First of all, notice that
the sensitivity is strongly affected by the intensity of the beam.  No
improvement on the existing bounds on these two parameters is achieved
after 5 years of data taking at nominal CNGS beam intensity, for any
of the considered value of $\delta_3$. Already with a doubled flux
intensity, some sensitivity to $\theta_{24}, \theta_{34}$ is
achievable. The sensitivity enhancement strongly depends on the value
of the CP-violating phase $\delta_3$, however.  For $\delta_3 = 0$,
OPERA can exclude a small part of the 99\% CL allowed region, only. 
On the other hand, for $\delta_3 = 90^\circ$ twice the nominal CNGS
flux suffices to put a bound on $\theta_{34} \leq 25^\circ$ for
$\theta_{24} \geq 4^\circ$ at 99\% CL. For maximal CP-violating
$\delta_3$, increasing further the CNGS flux can significantly
constrain the $(\theta_{24},\theta_{34})$ allowed parameter space.
Notice, eventually, the strong correlation between $\theta_{24}$ and
$\theta_{34}$ in the right panels of Fig.~\ref{fig:opera}. This is an
indication that the dominant channel that constrains these angles is
$\nu_\mu \to \nu_\tau$. As it can be seen in Eq.~(\ref{eq:pmutau}), 
the two angles always appear in combination, with an approximate
exchange symmetry $\theta_{24} \leftrightarrow \theta_{34}$.

\begin{figure}[t]
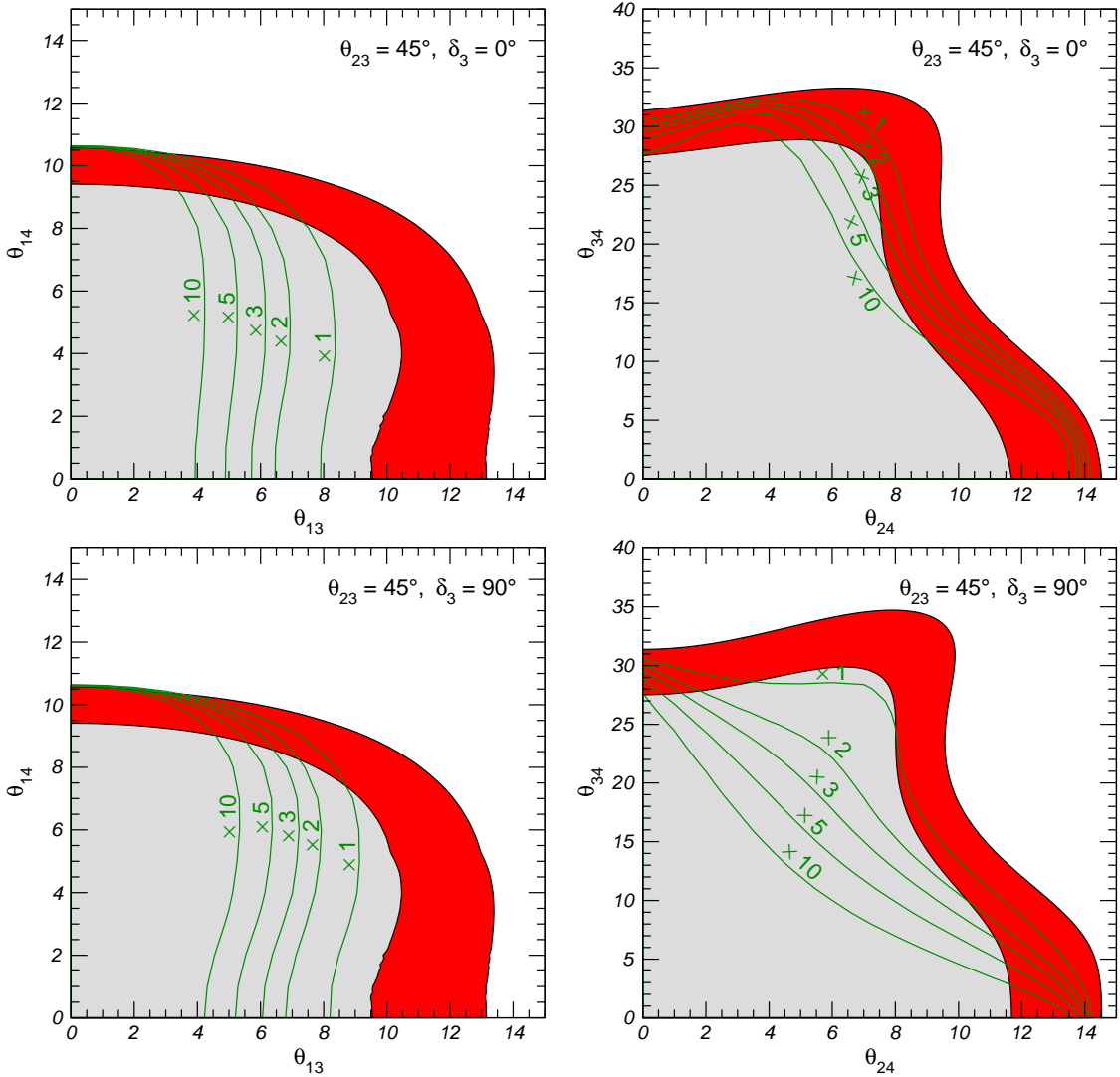
 \centering
    \includegraphics[width=0.95\textwidth]{fig.future-00.eps}
    \includegraphics[width=0.95\textwidth]{fig.future-90.eps}
    \caption{\label{fig:future}\sl%
      Sensitivity limit at 99\% CL in the $(\theta_{13},\,
      \theta_{14})$ plane (left) and in the $(\theta_{24},\,
      \theta_{34})$ plane (right) from the combined analysis of
      present data and a null result of the OPERA experiment, assuming
      1, 2, 3, 5 and 10 times the nominal intensity of $4.5 \times
      10^{19}$ pot/year. The coloured regions show the present bounds
      at 90\% and 99\% CL. We assume $\theta_{23} = 45^\circ$ and
      $\delta_3 = 0^\circ$ (top) or $\delta_3 = 90^\circ$ (bottom).}
\end{figure}

The allowed regions at 99\% CL in the $(\theta_{13},\,\theta_{14})$
plane (left) and in the $(\theta_{24},\,\theta_{34})$ plane (right)
from the combined analysis of present data and a null result of the
OPERA experiment after 5 years of data taking (assuming 1, 2, 3, 5 and
10 times the nominal CNGS intensity of $4.5 \times 10^{19}$ pot/year) 
are eventually shown in Fig.~\ref{fig:future}.  The coloured regions
refer to the present bounds at 90\% and 99\% CL, for $\theta_{23} =
45^\circ$ and $\delta_3 = 0^\circ$ (top) or $\delta_3 = 90^\circ$
(bottom). As it can be seen, the sensitivity of OPERA strongly benefits
from the complementary information on the neutrino parameters provided
by other experiments. In this case, even with the nominal beam
intensity the extension of the allowed regions is reduced by a
moderate but non-negligible amount.

\section{Conclusions}
\label{sec:concl}

The results of atmospheric, solar, accelerator and reactor neutrino
experiments show that flavour mixing occurs not only in the quark
sector, as it has been known for long, but also in the leptonic
sector. Experimental data well fit into a three-family scenario. The
existence of new ``sterile'' neutrino states with masses in the eV
range is not excluded, however, provided that their couplings with
active neutrinos are small enough. 

In this paper, we have tried to test the potential of the OPERA
experiment at the CNGS beam to improve the present bounds on the
parameters of the so-called four-neutrino models.  The model, in which
only one sterile neutrino is added to the three active ones
responsible for solar and atmospheric oscillations, is the minimal
extension of the standard three-family oscillation scenario.

We have determined the presently allowed regions for all
active-sterile mixing angles and studied the OPERA capability to
constrain them further using both the $\nu_\mu \to \nu_e$ and $\nu_\mu
\to \nu_\tau$ channels. We have performed our analysis using the OPERA
detector as a reference.  It can be extended including a detailed
simulation of the ICARUS detector at the CNGS beam.

Our conclusions are the following: if the OPERA detector is exposed to
the nominal CNGS beam intensity, a null result can improve a bit the
present bound on $\theta_{13}$, but not those on the active-sterile
mixing angles, $\theta_{14},\theta_{24}$ and $\theta_{34}$. If the
beam intensity is increased by a factor 2 or beyond, not only the
sensitivity to $\theta_{13}$ increases accordingly, but a significant
sensitivity to $\theta_{24}$ and $\theta_{34}$ is achievable. The
($\theta_{24},\theta_{34}$) sensitivity strongly depends on the value
of the CP-violating phase $\delta_3$, however, with stronger
sensitivity for values of $\delta_3$ approaching $\pi/2$. Only a
marginal improvement is achievable on the bound on $\theta_{14}$, that
should be constrained by high-intensitiy $\nu_e$ disappearance 
experiments. 

Notice that our results hold for any value of $\Dmq_\Sbl \geq
0.1~\eVq$, \textit{i.e.}\ in the region of $L/E$ for which
oscillations driven by this mass difference are effectively averaged.

\section*{Acknowledgements}

We acknowledge E.~Fern\'andez-Mart\'{\i}nez, P.~Hern\'andez,
J.~L\'opez-Pav\'on, M.~Sorel and P.~Strolin for useful discussions and
comments.  We thank T.~Schwetz for pointing out to us an error in the
first version of the paper and for useful comments on it. The work has
been partially supported by the E.U.\ through the BENE-CARE networking
activity MRTN-CT-2004-506395. A.D.\ received partial support from
CiCYT through the project FPA2006-05423.  M.M.\ received partial
support from CiCYT through the project FPA2006-01105 and the MCYT
through the \emph{Ram\'on y Cajal} program.  A.D.\ and M.M.\
acknowledge also financial support from the Comunidad Aut\'onoma de
Madrid through the project P-ESP-00346. D.M.\ would like to thank
CERN, where part of this work has been accomplished.

\end{document}